\documentstyle[12pt]{article}
\renewcommand{\baselinestretch}{1.5}
\input epsf

\let\a=\alpha \def\b{\bar{\a}} \let\g=\gamma \let\d=\delta
\let\e=\epsilon   
\let\th=\theta   \let\l=\lambda
\let\m=\mu \let\n=\nu  \def\p{\vec{p}} \def\q{\vec{q}}
\let\r=\rho \let\s=\sigma \def\t{\tilde{t}} \let\o=\omega 
   
\let\O=\Omega \let\S=\Sigma  \let\Th=\Theta \let\L=\Lambda
\let\G=\Gamma \let\D=\Delta  

\def\2{{1\over2}} \def\4{{1\over4}} \def\52{{5\over2}} \def\6{\partial }
\def\({\left(} \def\){\right)} \def\<{\langle } \def\>{\rangle }

\def\CV{{\cal V}}

\def\beg{\begin{equation}}
\def\begar{\begin{eqnarray}}
\def\ee{\end{equation}}
\def\ea{\end{eqnarray}}

\newcommand{\pref}[1]{(\ref{#1})}                
\newcommand{\fig}[1]{fig. \ref{#1}}              
\newcommand{\plabel}[1]{\label{#1}}              
\newcommand{\pcite}[1]{\cite{#1}}                
\newcommand{\pbib}[1]{\bibitem{#1}}              
\renewcommand{\Im}{\mbox{Im}}
\newcommand{\ps}{p\!\!\!/}                       

\begin{document}


\pagestyle{empty}

\topmargin-2cm

\hfill TUW-95-22

\vspace{.7cm}
\begin{center}
\Large
{\bf \renewcommand{\baselinestretch}{.5}
On the P-Wave Contributions to the Cross Sections
of $t\bar{t}$ and $\t\bar{\t}$ Near Threshold}\\[1cm]
\normalsize
Wolfgang M\"odritsch$^*$ \\[.6cm]
{\renewcommand{\baselinestretch}{1}
\small Institut f\"ur Theoretische Physik\\ Technische Universit\"at Wien\\
Wiedner Hauptstra\ss e 8-10/136, A-1040 Wien\\ Austria\\}
\end{center}
\vspace{.7cm}

\centerline{{\bf Abstract}}

\vspace{.7cm} \noindent
\parbox{16cm}
{\renewcommand{\baselinestretch}{1}
The approach used for the determination
of S-wave amplitudes containing the application of the nonrelativistic
approximation in the case of P-waves leads  to unphysical divergencies.
We show how to avoid the latter in calculations of contributions to the
cross section  near threshold in agreement with field theory.
This enables us to give quantitatively
reliable predictions for the forward-backward asymmetry and for the axial
contribution to the total cross section for the top-antitop system.
Also the cross section for the production
of stop-antistop near threshold is determined.

\baselineskip22pt }

\vspace{4cm}

\scriptsize
\_\hrulefill \hspace*{8cm} \\
Wien, August 1995\\
$^*$E-mail: wmoedrit@ecxph.tuwien.ac.at

\textheight21.5cm

\newpage
\pagestyle{plain}
\topmargin0cm

\normalsize
\section{Introduction}
The observation of the top quark has been reported recently by
two colaborations at FNAL \pcite{CDF,D0} (CDF and D0).
They quote masses of $176\pm8\pm10GeV$ and
$199^{+19}_{-21}\pm22 GeV$ respectively . Electroweak data
from LEP \pcite{LEP} and the predicted Standard Model cross section
indicate a slightly lower top mass around $170GeV$.

Thus it seems to be established by now that the top quark is very
heavy, giving it an
extraordinary status in the particle spectrum discovered so far.
A main characteristic of the top quark compared to
other heavy quarks is its large decay width due to $t\to W^+b$ of the
order of $1GeV$.
In the present paper we will investigate properties of the top which are
relevant in a future $e^+e^-$ collider (or perhaps $\m^+ \m^-$).

Such a machine would be the ideal place to
investigate in detail its properties. Especially the
threshold region could give us independent determinations of the mass
and width of the top quark as well as an improved determination of
the strong coupling constant.
Whereas for lighter quarks the $q\bar{q}$ bound state shows
individual energy levels (as in charmonium and bottonium) the large
electroweak decay width of the top smears out individual levels. On
the other hand it makes QCD perturbation theory more reliable
\pcite{Khoze} because the $t$ decays fast enough on the times scale of
strong interaction to prevent hadronisation and thus essential
contributions of inherent nonperturbative character.

This work is organized as follows. In sect. 2 we fix the notation for
cross sections and summarize the Green function approach. Section 3
is devoted to the calculation of the forward-backward asymmetry
originating in the interference of S- and P-waves.
P-wave Green functions are needed exclusively for the calculation of the
total cross section for the stop-antistop production as well as
for the axial contribution to the cross section for $t\bar{t}$.
Both are calculated in chapter 4.
Finally in chapter 5 we draw our conclusions.

\section{Calculation of cross sections near threshold}

The nowadays quite standard Green function approach
\pcite{Sumino,Jeza,Peskin} relies on the
observation that particles produced near threshold move relatively slow
and thus one approximates the exact four point function by a
nonrelativistic Green function. This Green function fulfills the equation
\beg \plabel{Schroed}
\{ -\frac{\Delta}{m} + V(r) -E- i\G \} \tilde{G}(\vec{r},\vec{r}\,') =
         -\d(\vec{r}-\vec{r}\,')
\ee
where $V(r)$ is a perturbative or QCD-motivated potential. The width
$\G$ has been included to treat an unstable particle.
In general the total cross section for $t\bar{t}$ production
is given by ( a color factor 3 already included, $X=A,V$ )
\begar
  \s &=& \s_V +\s_A \plabel{stot}\\
 \s_X &=& \frac{72}{P^2} \pi  [ c_X + d_X \r]  \Im G^{XX}(P)
\s_{\m \bar{\m}},
   \plabel{stotx}
\ea
whereas for scalar constituents, like $\t\bar{\t}$ (stop-antistop)
from an eventual supersymmetric generalization
only one contribution occurs
\beg
\s_s =  \frac{72}{P^2}  [c_s + d_s  \r]  \Im G_s(P) \s_{\m\bar{\m}}
\ee
with
\begar
  \begin{array}{ll}
   c_X := (a^{(X)})^2  + (b^{(X)})^2 & c_s = a_s^2  + b_s^2 \\
   d_X := 2 a^{(X)} b^{(X)} & d_s = 2 a_s b_s
  \end{array}
\ea
and
\beg
 \s_{\m\bar{\m}} = \frac{4\pi \a_{QED}^2}{3 P^2}, \quad P=2m+E.
\ee
The Standard Model (SM) values of $a^{(X)},b^{(X)}$ as well as the
values of
the minimally supersymmetric extended Standard Model (MSSM)
$a_s,b_s$ are given by
\begar
 a^{(V)} &=& q_t - \frac{v_t v_e P^2}{4 s^2 c^2 z}, \qquad
       b^{(V)}=\frac{v_t a_e P^2}{4 s^2 c^2 z}, \plabel{av} \\
 a^{(A)} &=& \frac{a_t v_e P^2}{4 s^2 c^2 z}, \qquad b^{(A)}=-\frac{a_t
 a_e P^2}{4 s^2 c^2 z},\plabel{aa}
\ea
and
\begar
a_s &=& -q_t + \frac{\tilde{Q}_{Z} P^2}{2 s c z}, \qquad
       b_s=-\frac{\tilde{Q}_{Z} a_e P^2}{2 s c z},
\ea
where we used the abbreviations
\begar
z &=& P^2 -M_Z^2+i M_Z \G_Z, \quad s=\sin \th_W, \quad c = \cos \th_W
\plabel{z} \\
v_f &=& T_3^ f - 2 Q_f  \sin^2 \th_W, \quad a_f = T_3^f
\ea
Here $T_3^f$ is the eigenvalue of the diagonal SU(2) generator for
the fermion $f$ (e.g. $t$ or $e$)with charge $Q_f$.
For the supersymmetric charges we use \pcite{Been}
\begar
 \tilde{Q}_{\g} &=& - q_t, \\
 \tilde{Q}_{Z} &=& (\cos^2 \th_t- 2 q_t \sin^2 \th_W)/(2 s c).
\ea

The leading contributions to \pref{stot} arises from the vector coupling
since it produces $t\bar{t}$ pairs  with angular momentum zero (S-waves).
In the fermionic case P-waves are produced by the axial current
and represent higher order correction.
They contribute to $O(\a_s^2)$ to the total cross section and lead to
a forward backward asymmetry \pcite{Sumino2} by interfering with the vector
contribution. On the contrary,
for scalar particles P-waves represent the dominant contribution
near threshold. The four point functions $G^{XX}(P),G_s(P)$ are in the
nonrelativistic approximation replaced by
\begar
 \Im G^{VV}(P) \to \frac{1}{m^2} \Im \tilde{G}(0,0) \plabel{nrlimv} \\
 \Im G^{AA}(P) \to \frac{4}{m^2} \Im  \frac{\6}{\6 x_3}\frac{\6}{\6 y_3}
           \tilde{G}(\vec{x},\vec{y}) \big|_{|\vec{x}|\to0,|\vec{y}| \to
            0} \plabel{nrlima} \\
 \Im G_s(P) \to \frac{1}{m^2} \Im  \frac{\6}{\6 x_3}\frac{\6}{\6 y_3}
           \tilde{G}(\vec{x},\vec{y}) \big|_{|\vec{x}|\to0,|\vec{y}| \to
            0} \plabel{nrlims}
 \ea
While $\Im \tilde{G}(0,0)$ in eq. \pref{nrlimv} is well behaved, the r.h.s
in eqs. \pref{nrlima} and \pref{nrlims} are linearly divergent.

Befor tackling this problem, let us consider the calculation of the
nonrelativistic Green function. Due to its symmetry
and the requirement of regularity at $r \to \infty$ and
at the origin, the general solution of eq. \pref{Schroed} can be written as
\begar
\tilde{G}(\vec{r},\vec{r}\,') &=& \sum_{l=0}^{\infty} g_l(r,r')
\sum_{m=-l}^l Y_{lm}^*(\O')
             Y_{lm}(\O) \plabel{gsum} \\
      g_l(r,r')  &=& \frac{ g_<(r_<) g_>(r_>)}{r_< r_>}, \qquad
       r_> = \left\{ \begin{array}{c} r: r> r' \\ r': r'> r \end{array}
       \right. \qquad
       r_< = \left\{ \begin{array}{c} r: r< r' \\ r': r'< r \end{array}
       \right.
\ea
where $ g_<(r_<)$ and $g_>(r_>)$ are regular solutions of the homogeneous
equation
\beg \plabel{gl}
 \{ \frac{\6^2}{\6 r^2} - m(E+i\G-V(r)) -  \frac{l(l+1)}{r^2} \} g(r) = 0
\ee
 at $r=0$ or
$r \to \infty$, respectively. The actual behavior of the solution
$ g_<(r_<)$ (and also of the irregular one) for $r \to 0$  depends on
the potential and on $l$. In this work we use a one loop renormalization
group improved potetial, treating four flavors as massles ($n_f=4$).
The bottom quark contribtion is included in a pure perturbative manner.
\begar
 V(r) = \frac{\a}{r [1- \frac{(33-2 n_f) \a}{8 \pi}( \g+ \ln \m r)]}
       + \frac{\a^2}{4\pi r}[Ei(-r m_b e^{\frac{5}{6}}- \frac{5}{6}+
        \2 \ln (\frac{\m^2}{m_b^2} + e^{\frac{5}{3}} )]
\ea
The color factor $4/3$ is absorbed in the definition of $\a$:
\beg \plabel{alpha}
 \a := \frac{4}{3} \a_s = \frac{g^2}{3 \pi}
\ee
In principle all other parts of the general potential derived in
\pcite{KM1} on purely field theoretic grounds can
be included as well.

Due to the presence of the width $i\G$ the
behavior of $ g_>(r)$ for $r \to \infty$ is given by
\begar
  \lim_{r \to \infty} g_>(r) \propto e^{-a_-r}
\ea
with
\beg
 a_- := \sqrt{\frac{m}{2}} \sqrt{ - E + \sqrt{E^2+\G^2}}.
\ee
This exponential damping behavior is the origin of the infrared cut
off mentioned already
in \pcite{Khoze}. Consider for simplicity the case $E=0$. Then we can
write
\begar \plabel{damp}
a_-^{-1} = \sqrt{\frac{2}{m \G}} =  \sqrt{\frac{E_B}{\G}} r_B
\ea
where $E_B=m \a^2/2$ and $r_B$ are the Bohr energy and the Bohr radius of
the system, respectively . This formula clarifies
that if the width $\G$ is approximately equal to $E_B$ the form
of the potential is only "tested" up to the order of magnitude of the
Bohr radius. The situation is improved for $E< -\G$ since here we
have
$$ a_-^{-1} \le \sqrt{\frac{1}{m|E|}}, \quad  E< -\G.$$
This relation also holds also for small $\G$ and thus the Bohr radius
again becomes the relevant scale.

The matching condition for $g_<$ and $g_>$ is provided by the $\d$
distribution in eq. \pref{Schroed}:
\beg \plabel{Wronski}
 -m =  g_<(r) g_>'(r)- g'_<(r) g_>(r)
\ee
Numerically it is possible to obtain the regular solution at the origin
directly by imposing suitable boundary conditions.
In the following we will need both S- and P-wave Green functions.
For the singular solution of the former
we use the following method \pcite{Sumino,Peskin}.
Suppose you have two solutions:
$ g_<(r)$ as above and $u_2(r)$ a solution determined by an arbitrary
boundary condition. Then the solution $g_>$ is given by
\beg
 g_>(r) = c[ u_2(r) + B  g_<(r)  ]
\ee
with
\beg B = -\lim_{r \to \infty} \frac{u_2(r)}{g_<(r)}.
\ee
The constant $c$ can be determined from eq. \pref{Wronski}. For $l=0$ and
a Coulomb-like potential at the origin we arrive with the boundary
conditions
\begar
 g_<(0) &=&0 \\
 g_<'(0) &=& 1 \nonumber
\ea
and arbitrary ones for $u_2$ at the result
\begar
\tilde{G}_{l=0}(\vec{r},0) &=& \frac{m}{4 \pi w} \frac{u_2(r) +
B  g_<(r)}{r},   \plabel{sGr}\\
 w &=& u_2 g_<' - g_< u_2'.
\ea
According to eq. \pref{stotx} and \pref{nrlimv} the total cross
section is proportional to
$\Im \tilde{G}(0,0)$. With this knowledge it is possible to calculate
the total
cross section near threshold for any given potential $V(r)$. We have
written a numerical routine for MATHEMATICA$^{TM}$ which has been
checked to give the correct answer for the known Green function of a
purely Coulombic potential.

Before presenting some numerical studies, a short remark on the coupling
constant to be used seems in order. It is customary to take
the strong coupling constant $\a_s$
in the $\overline{\mbox{MS}}$ scheme. Especially experimental determinations
are always given in terms of $\a_s^{\overline{MS}}(M_Z)$. But in bound state
calculations it is natural to use an $\a_s$ defined differently.
Clearly it should be possible to relate the two schemes.
To avoid complications from heavy fermion masses we treat 4 flavors as
massless. The difference in the bottom quark contribution can be expected
to be smaller than the experimental uncertainties in $\a_s$. Therefore
we ignore it in the following consideration.
The quark-antiquark potential  in the $\overline{\mbox{MS}}$ scheme can be
extracted e.g. from
\pcite{Billoire} to $O(\a_s)$:
\beg
 \CV = - \frac{4\pi \a}{\q\,^2} \left[ 1 - \a (3 \pi \b_0 \ln
          \frac{\q\,^2}{\m^2} + \frac{31}{16 \pi} -  \frac{10 n_f}{16
          \pi})  \right]
\ee
Comparison with the QCD potential \pcite{KM1} for $n_f$ light flavors gives
\beg
\a^{BS}=\a_{\overline{MS}}\left(1- \frac{\a_{\overline{MS}}}{16 \pi}
(31-\frac{10 n_f}{3} )\right)
\ee
where $BS$ denotes our bound state scheme used in \pcite{KM1}.
Numerically this means that
our $\a$ is slightly smaller than usual which is advantageous for our
perturbative calculation. For $m_t = 180, \a_s^{\overline{MS}}(M_Z)
=.117\pm0.05$ get for a renormalization at $\m = 1/r_B(\m)$
\begar
 \m &=& 18.5\pm1 GeV \\
 \a^{\overline{MS}}(\m) &=& 0.22 \pm 0.01 \\
 \a(\m)&=&0.19 \pm0.01
\ea
Remember that the above values for $\a,\a^{\overline{MS}}$ differ by a
factor $4/3$ from $\a_s$.

\section{Forward-Backward Asymmetry}

The formalism used in the last section to investigate
the total cross section for $t\bar{t}$ production
addressed
a very suitable observable to determine the mass of the top quark. But
since the total cross section at threshold is also significantly
influenced by other quantities like the strong coupling constant and the
top decay-width, it is necessary to have some independent observables.
It was first proposed in \pcite{Sumino2} that the forward backward (FB)
asymmetry would provide such a quantity.
It has the additional advantage that it is independent of the absolute
normalization of the cross section measurement. Thus it can be
determined experimentally with high accuracy.
However, treating this problem
in a nonrelativistic context, one encounters unphysical divergences which
made necessary the introduction of an unnatural cut-off. In
this section we will show a systematic way how to avoid such
divergencies.

The FB asymmetry is defined as the relative difference of the
cross section for particles produced in the forward and in the backward
direction, with respect to the direction of the $e^-$ beam.

\begar
 A_{FB} := \frac{\s_{FB}}{\s_{tot}} = \frac{ \int_{0}^{\frac{\pi}{2}} d\th
\frac{d\s}{d\th}- \int_{\frac{\pi}{2}}^{\pi} d\th \frac{d\s}{d\th}}
{\s_{tot}}
\ea

Since the top quarks will be identified by their main decay products
$W$ and $b$ let us consider the cross section for the process
$e^+ e^- \to t\bar{t} \to b W^+ \bar{b} W^-$.
\begar
 d\s_{e^+ e^- \to bW^+\bar{b}W^-} = \frac{1}{2 P^2} |M_A+M_V|^2 d\Phi_{W^+b}
 d\Phi_{W^-\bar{b}} (2\pi)^4 \d(t_1+t_2-k_1-k_2-b_1-b_2)
\ea
with the phase space factor
\begar
d\Phi_{W^+b} =  \frac{d^4b}{(2 \pi)^4}\frac{d^4k}{(2 \pi)^4}  (2\pi)
\Th(k_{10}) \d(k^2-M^2) (2\pi)
      \Th(b_{10}) \d(b_1^2)  \plabel{dPhi}
\ea
and a similar one for the decay of the antitop.
Production of $W^+bW^-\bar{b}$ via other channels can be treated as in
\cite{Balle}.

\begin{figure}
\begin{center}
\leavevmode
\epsfxsize=16cm
\epsfbox{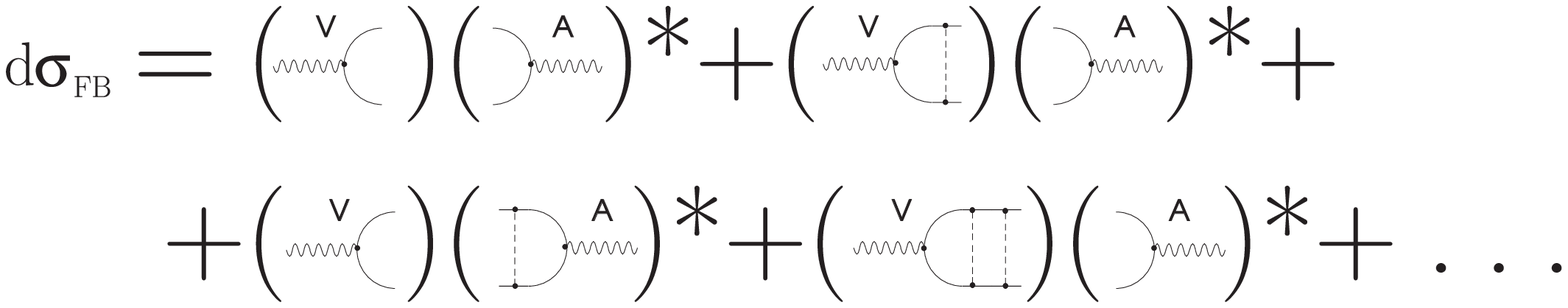}\\
Fig. \refstepcounter{figure} \label{fbf1} \thefigure
\end{center} \end{figure}

The FB asymmetry originates in the interference terms of the
vector coupling ($M_V$) and the axial coupling ($M_A$).
This is depicted in figure \pref{fbf1} where V indicates a
vector and A an axial vector coupling, the star denotes complex
conjugation.
The matrix element $M$ is essentially the vertex $\g t \bar{t}$ and $Z t
\bar{t}$. Near threshold the perturbative treatment of the Coulomb
interaction is no longer valid. Instead one has to include the whole rung
of Coulomb interactions, and even worse, it is necessary to include also the
leading logarithmic contributions from the gluonic self energy
corrections. We can sum the relevant set of graphs by means of the equation
\beg
 \G^{\m} = \g^{\m} -i KD\G^{\m} \plabel{vertgl}
\ee
for the vertex function $\G^{\m}$. $D$ contains the unconnected
fermion propagators and $K$ represents the sum of all 2pi graphs.
Comparison with the BS-Equation
for the Feynman amplitudes
\beg
 iG=-D+DKG
\ee
shows that eq. \pref{vertgl} has the solution
\begar
 \G_V ^{\mu} &=& -i D^{-1} G \g^{\mu} \\
 \G_A ^{\mu} &=& -i D^{-1} G \g_5 \g^{\mu},
\ea
where it is understood, that the two legs of $G$ on the right hand side
are connected with the $\g$-matrices and a momentum integration is performed.

We can split the matrix element into
\begar
M_A^i &=& m_1 m_2 D \G_A ^{i} \\
M_V^i &=& m_1 m_2 D \G_V ^{i} \nonumber
\ea
where
\begar
m_1 &=& \e_{\m}^{(\s)}(k_1) \bar{u}_{\l}(b_1) (-i \frac{e}{\sqrt{2} \sin
      \theta_W} V_{33}^* \g^{\m} P_-) \\
m_2 &=& \e_{\n}^{*(\s')}(k_2)  (-i \frac{e}{\sqrt{2} \sin
      \theta_W} V_{33} \g^{\n} P_-) v_{\l'}(b_2).
\ea
Near threshold the propagators in $D$ and the Green function $G$ should be
dominated by small momenta and therefore it should be possible to replace
them by their nonrelativistic approximations. However, for $D$ the
approximation to $O(\a)$ is needed for the axial vertex, since the zero
order propagators would give zero due to $\l^-\g_5 \g^i \l^+ =0$.
\begar
 D &\to & S^+_{nr} \otimes S^-_{nr} \\
        S^{\pm}_{nr} &=& (\l^{\pm} - \frac{\vec{p}\vec{\g}}{2m})
           \frac{1}{\frac{E}{2} \pm p_0 -\frac{\p\,^2}{2m}+i\frac{\G}{2}}
\ea
This gives rise to the effective nonrelativistic axial vertex
\beg
 \g_5 \g^i \to \frac{p^k}{2m} [\g^i,\g^k] \g_5 \l^- \plabel{AVnr}.
\ee
We note already here that the term $\frac{\vec{p}\vec{\g}}{2m}$
introduces an additional power in $\p$ which is not present in the
relativistic propagator. Instead the factor $\frac{p^k}{2m}$ in
\pref{AVnr} would be replaced by $\frac{p^k}{2 E_p}$ which is finite for
$p \to \infty$. Thus a divergence of \pref{AVnr} at $|\p| \to \infty$ is
really an artifact of the nonrelativistic approximation.

Using the fact that the heavy quarks are on shell up to $O(\a^2)$
we can write
\begar
\int d\Phi_{W^+b} \sum_{\s,\l} |m_1|^2 = - 2 \Im \S(m^2) \l^+ + O(\a^2)
      \plabel{imsig}
\ea
in  agreement with \pcite{topdec}.

{}From the modulus squared of the propagators we obtain a factor
\begar
\int \frac{dp_0}{2\pi} |\frac{1}{p_0^2-\o^2}|^2 &=& \frac{2}{\G |2
 \o|^2}, \\
 \o &=& \frac{1}{2m}(\p\,^2-mE-im\G)
\ea

Collecting everything from above, performing the trace in a frame where
the leptons move along the z-axis we get for the FB-asymmetry

\begar
\s_{FB,\L} &=& (c_{AV} + d_{AV} \r) \frac{18 \G}{ \pi^2 P^2}
\int_0^{\L} p^2 dp d \O_{FB} Re[ G^*(\p) \int
           \frac{d^3q}{(2 \pi)^3} G(\p,\q) \frac{q^{(3)}}{m} ]
           \s_{\m\bar{\mu}} \plabel{sfbnr}
\ea
where
\begar
       c_{AV} &=& a^{(A)} b^{(V)} +a^{(V)} b^{(A)} \\
       d_{AV} &=& a^{(A)} a^{(V)}+b^{(A)} b^{(V)} \\
\int d \O_{FB} &=& \int_0^{2\pi} d\varphi ( \int_0^{\frac{\pi}{2}}
 d\th  - \int_{\frac{\pi}{2}}^{\pi} d\th)\sin\th
\ea
The SM values for $a^{(X)}, b^{(X)}$ are given in \pref{av} and \pref{aa}.
We introduced an intermediate cut-off $\L$ in \pref{sfbnr}
since the $p$ integration
is logarithmically divergent. It will be possible to
remove it in the following.
The Green function $G(\p)$ in \pref{sfbnr} is defined by
\begar
 G^*(\p) &=& \int \frac{d^3q}{(2 \pi)^3} G^*(\p,\q) =
        \frac{4\pi}{p} \int_0^{\infty} dr r \sin pr \tilde{G}_{l=0}(r,0).
\plabel{Gs}
\ea
$\tilde{G}_{l=0}(r,r')$ denotes the S-wave Green function in
configuration space, eq. \pref{sGr}. Let us now investigate the term
\begar
 \int \frac{d^3q}{(2 \pi)^3} G(\p,\q) \frac{q^{(3)}}{m} =
 \frac{-i}{m} \int d^3x e^{-i\p\vec{x}} \frac{\6}{\6 y_3}
 \tilde{G}(\vec{x},\vec{y}) \big|_{\vec{y}=0} \plabel{Gp1}
\ea
Using the representation \pref{gsum} one can show that only P-waves
($l=1,m=0$) contribute to the sum. Thus we need to evaluate eq. \pref{gl}
for $l=1$. The regular and singular solutions behave at the origin as
\begar
 g_<^{l=1}(r) &\to& c_{\scriptscriptstyle <} r^2,  \\
 g_>^{l=1}(r) &\to& c_{\scriptscriptstyle >} \frac{1}{r}, \nonumber
\ea
respectively. We now construct a singular solution with
$c_{\scriptscriptstyle >}=1$ out
of two arbitrary solutions $u_{p1}$ and $u_{p2}$ :
\begar
 g_>^{l=1}(r) &=&  a_p[u_{p2}(r) - B_p u_{p1}(r)]
\ea
by means of
\beg
 a_p := \lim_{r \to 0} r^{-1}[ u_{p2}(r) - B_p u_{p1}(r)]^{-1}.
\ee
$B_p$ is determined by the requirement of regularity at infinity
\beg
 B_p = \lim_{r\to \infty} \frac{u_{p2}(r)}{u_{p1}(r)}.
\ee
Then the condition \pref{Wronski} demands $c_{\scriptscriptstyle
<}=m/3$. Performing the
differentation in the $z$-direction in the limit $y \to 0$ in \pref{Gp1}
leads to
\begar
 \int \frac{d^3q}{(2 \pi)^3} G(\p,\q) \frac{q^{(3)}}{m} &=&
 \frac{\cos \th}{p^2} \int_0^{\infty} dr ( \frac{\sin pr}{r} - p \cos pr)
         g_>^{l=1}(r). \plabel{Gp2}
\ea
While the expressions \pref{Gs} and \pref{Gp2} are well defined,
it has been observed in \pcite{Sumino2} and \pcite{jeza2}
that \pref{sfbnr} is logarithmically divergent. Thus cutoffs ($\L$) have
been introduced to make numerical predictions possible. Since in our
present case the divergence is only logarithmical, this leads to
phenomenologically reasonable results.
But the explanations
given in these references are different and so are the cut-offs.
It may, however, happen that for the comparison with experimental data
with some cuts a reintroduction of some kind of cut-off will be necessary
\footnote{The author is grateful to M.Je\.zabek for this information},
but we feel saver by first giving a satisfactory theoretically prediction
and leaving this possibility open to the experimentalists.

In any case, the introduction of these cut-offs leads to some numerical
uncertainty in the result, and
is clearly unsatisfactory from the theoretical point of view. Furthermore
we will see in the next section, that a similar UV-divergence occurs
in the context of the stop--anti-stop production. Since this divergence
is linear, a better understanding of the origin of this kind of divergencies
is necessary to give a quantitative prediction.

The key point of our solution of the problem is to go back to the
perturbative sum of the graphs for $\s_{FB}$. This is illustrated in
figure \pref{fbf1}.

Consider now the leading (tree) contribution in the nonrelativistic
approximation (e.g. the first graph in \fig{fbf1}). In this case
both S- and P-wave Green functions
are replaced by
\beg
 G(\p,\p\,') = \frac{(2\pi)^{3} \d(\p-\p\,')}{\frac{\p^2}{m} -E -i\G}.
\ee
This leads to the logarithmically divergent expression (for $\L \to
\infty$)
\begar
 \s_{FB,\L}^{(1),nr} = \frac{9 \G}{2 m^3} \int d\O_{FB} \int_0^{\L} dp
        \frac{p^3 \cos \th}{|\frac{\p^2}{m} -E -i\G|^2} \plabel{sfb1nr}.
\ea
Since the remaining graphs in \fig{fbf1} by power
counting can be shown  to give finite results we
conclude that the nonrelativistic approximation was not valid in the tree
graph due to the extra power in $p$ from the axial vertex. Furthermore
since \pref{sfb1nr} is the only divergent contribution in the
(infinite, but assumed to be convergent) sum in \fig{fbf1} representing $G$
we conclude that
the divergent part of \pref{sfbnr} is entirely contained in the first
(free) contribution \pref{sfb1nr}. We could now
remove this divergence by a replacement $p/m \to p/E_p$, as indicated
above. This, however, does not respect another qualitative difference
between the exact tree contribution and its nonrelativistic approximation.
Namely in the relativistic calculation the phase space is cut off
when the momentum squared of one quark falls below the invariant mass
of the decay products \pcite{Sumino}.

Therefore, we return, instead, to the relativistic expression for the tree
contribution. Since we are dealing with a decaying particle, we will
have to include the self energy contribution due to the decay in the
propagator. This will in general lead to gauge dependent results,
but we can use the constant on-shell width to obtain the leading
contribution in the weak coupling \pcite{Velt}.

This makes the quantities we are considering finite (for $\G\to0$) and
gauge independent.
The remaining higher order contributions (e.g. gauge dependent if one
considers
only one diagram) can be calculated
perturbatively. Therefore, to leading order it seems reasonable to take the
constant width approximation for the top propagator.

\begin{figure} \begin{center}
\leavevmode
\epsfxsize=7cm
\epsfbox{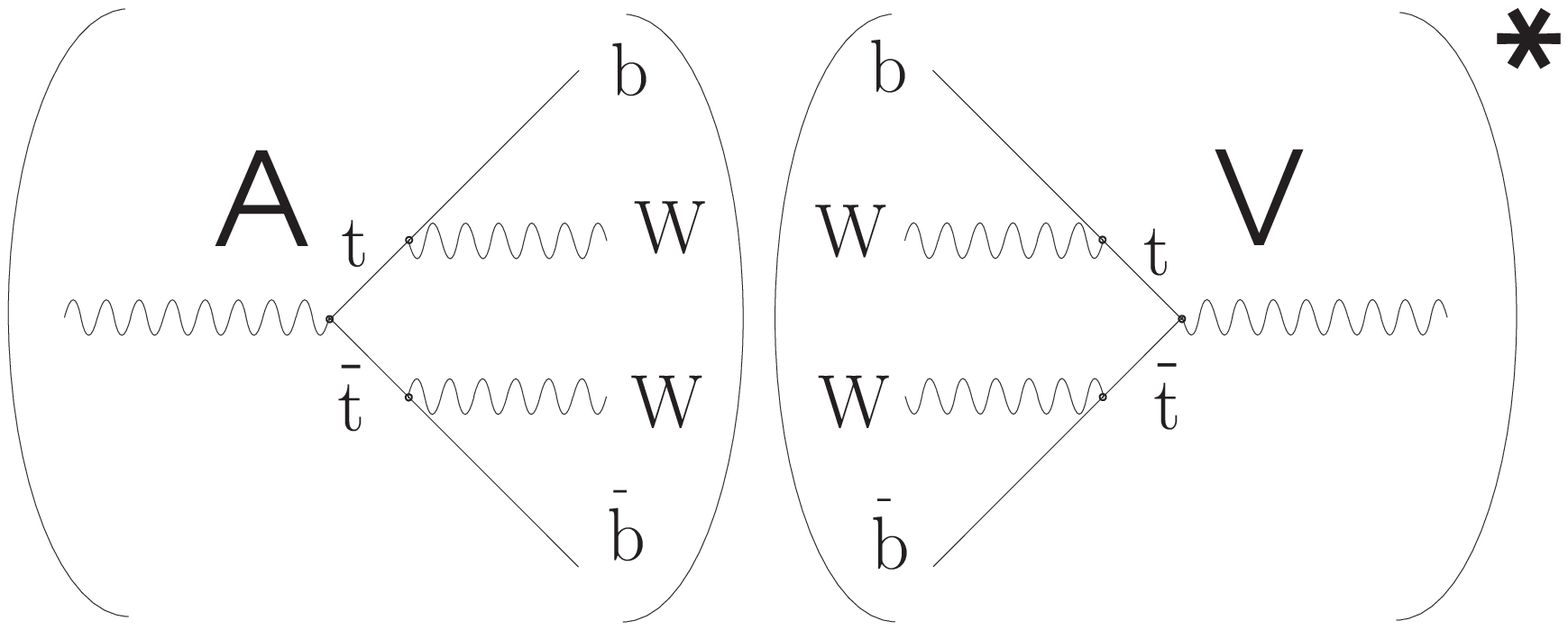}\\
Fig. \refstepcounter{figure} \label{reltree} \thefigure
\end{center} \end{figure}

Since we are only investigating processes
with $t\bar{t}$ intermediate states which will give the main contribution
to the cross sections, we focus on the graph shown in \fig{reltree}.
A further advantage of this method is that now
the other - non resonant - graphs for the process $e^+e^- \to
W^+bW^-\bar{b}$ to leading order in the weak coupling \pcite{Balle}
need only to be added to yield the background contribution.

Performing the straightforward calculation for the relativistic tree
contribution with a constant, but non-zero width $\G$, we arrive at
\beg
\s_{FB}^{(1)} = \frac{18}{P^2} (c_{AV} +c_{AV} \r) \int d\m_1^2 \int
d\m_2^2  \D(\m_1^2) \D(\m_2^2) \left(1-\frac{(\mu_1+\m_2)^2}{P^2}
        \right) \left(1-\frac{(\mu_1-\m_2)^2}{P^2} \right)
\ee
where
\beg
\D(p^2) = \frac{m \G}{\pi[(p^2-m^2)^2+m^2\G^2]}
\ee

This could have been also obtained by replacing the fermion propagator by
\beg
 S = \int d\mu^2 \frac{\G}{\pi[ (m^2-\mu^2)^2+m^2 \G^2 ]}
         \frac{(\ps+\m)}{p^2-\m^2+i\e}.
\ee
and cutting it to effectively replace the  phase space element for a
stable quark

$$ d\Phi_{stable} = \frac{d^4p}{(2\pi)^3} \Th(p_0) \d(p^2-m^2)$$ by
\begar
d\Phi_{unstable} &=& \frac{d^4p}{(2\pi)^3} \Th(p_0) \D(p^2).
\ea

A transparent summary of our approach may be formulated like this:
To obtain  finite,
gauge independent results to the desired order we simply propose to
replace the nonrelativistic tree contribution by the relativistic one
and leave the (finite) rest unchanged:
\beg
 \s_{FB}= \s_{FB}^{(1)}+ \lim_{\L \to \infty}
   (\s_{FB,\L}-\s_{FB,\L}^{(1),nr})
\ee
In \fig{fb1} the numerical results are shown for the different
contributions to $\s_{FB}$ for $\L=300$ GeV. This can be estimated to
give a result for $ \lim_{\L \to \infty} (\s_{FB,\L}-\s_{FB,\L}^{(1),nr})$
lying only 2\% below the final answer.
The results of the purely nonrelativistic calculations are
compared to our approach in \fig{fb2} for $m_t=180$ and an electron
polarization of 0.6. We also included the hard corrections as given in
\pcite{jeza2,ph}.

We believe that our approach has the decisive
advantage that it can be based upon
to the {\it original} set of graphs to be considered
and it is clear which graph has been
calculated to which accuracy. Thus at least in principle
a systematic improvement
is possible. The present approach should also be applicable to the
$O(\a_s)$ final state corrections calculated in \pcite{Sumith}.
\begin{figure}
\begin{center}
\leavevmode
\epsfxsize10cm
\epsfbox{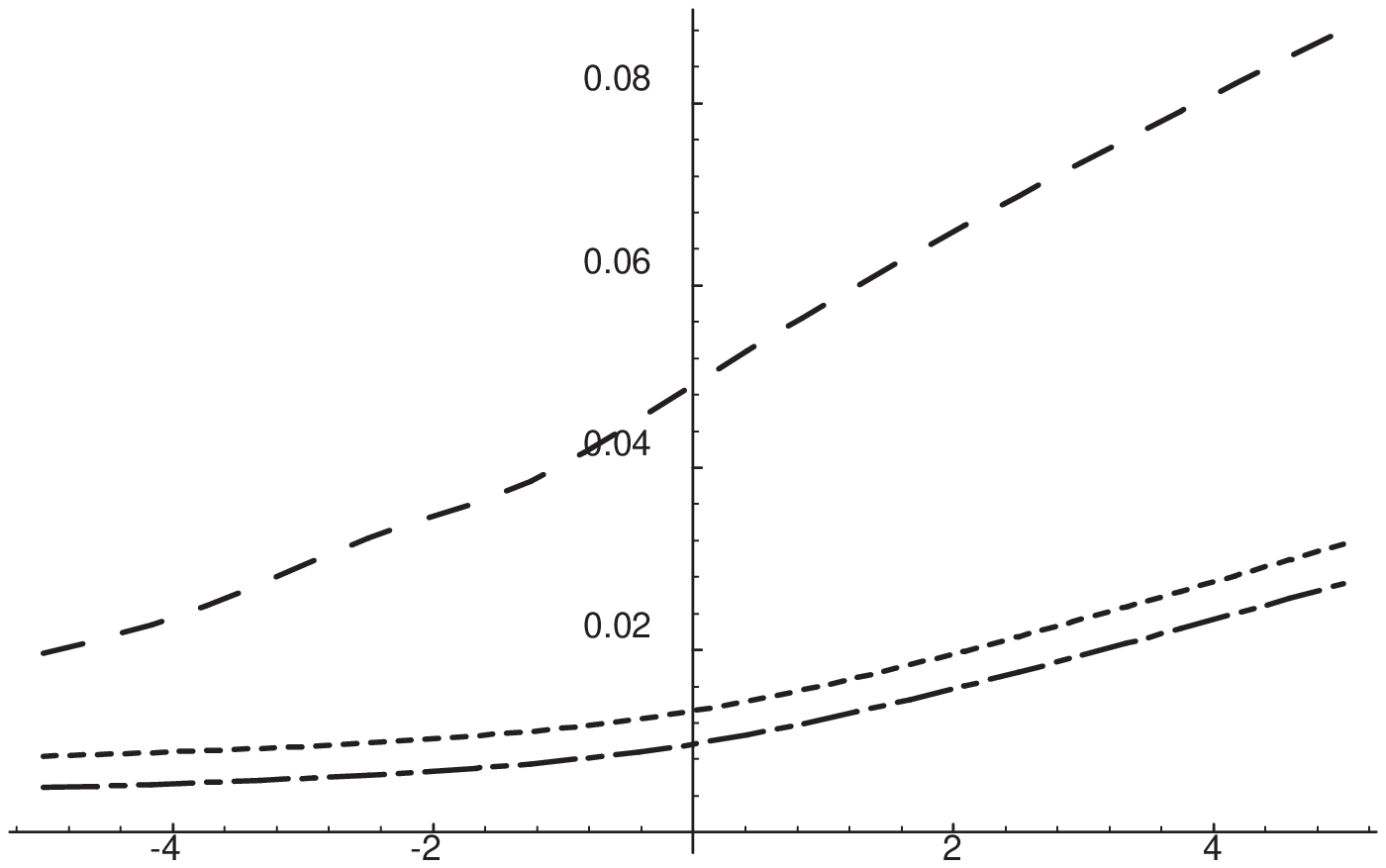}\\
Fig. \refstepcounter{figure} \plabel{fb1} \thefigure:
Different Contributions to $\s_{FB}$\\
dashed: $\s_{FB,\L}$, dotted:$\s_{FB,\L}^{(1),nr}$,
dashed-dotted: $\s_{FB}^{(1)}$
\end{center}
 \unitlength1mm
 \begin{picture}(0,0) \put(77,93){$\frac{\s}{\s_{\m\bar{\m}}}$}
   \put(130,28){E/GeV} \end{picture}
\end{figure}
\begin{figure}
\begin{center}
\leavevmode
\epsfxsize10cm
\epsfbox{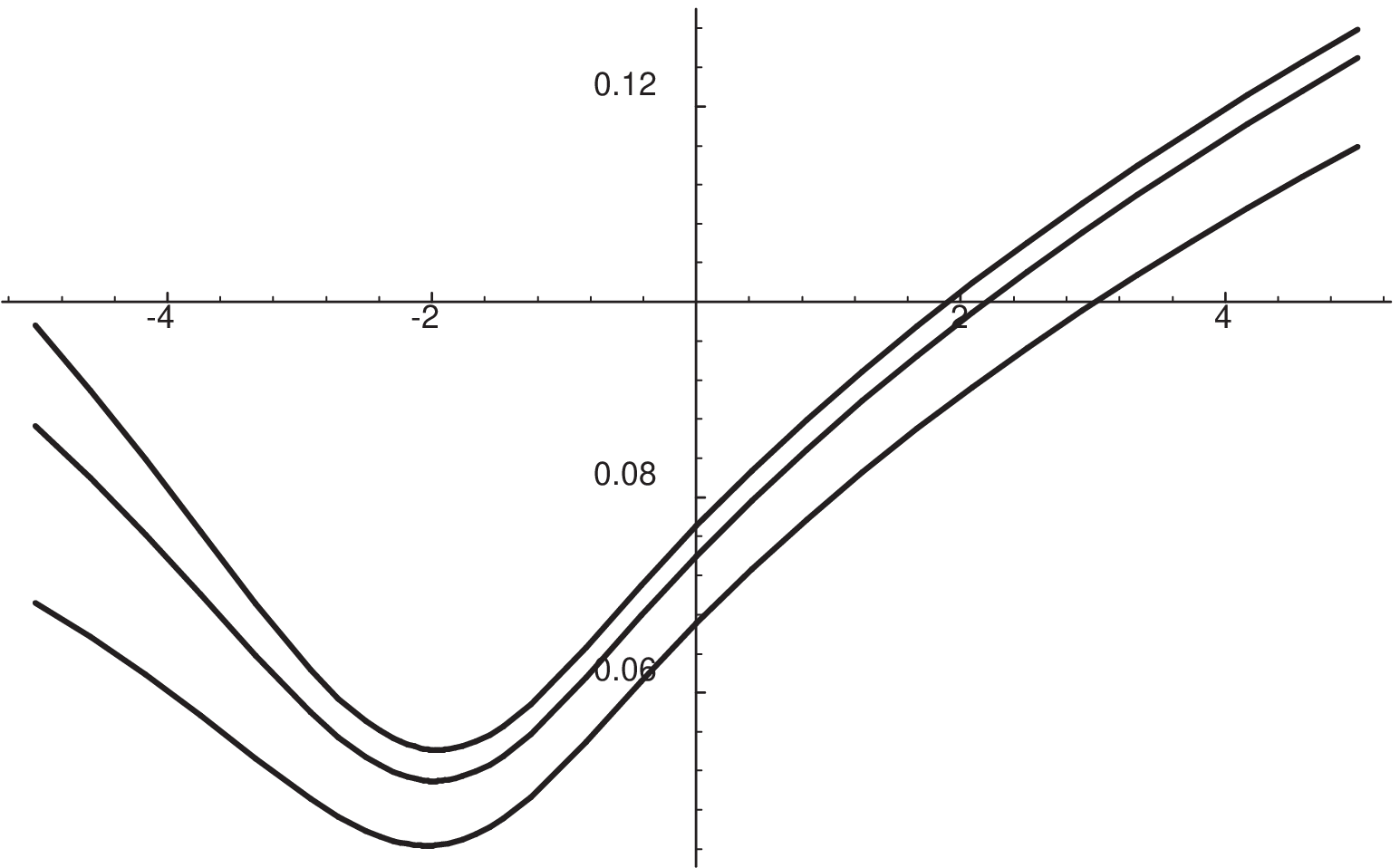}\\
Fig. \refstepcounter{figure} \plabel{fb2}\thefigure:
Forward-backward asymmetry $A_{FB}$;\\
curves from top to bottom: present work, ref.\pcite{Sumino2},
ref.\pcite{jeza2}
\end{center}
\unitlength1mm
\begin{picture}(0,0) \put(77,92){$A_{FB}$}
\put(130,66){E/GeV}
\end{picture}
 \end{figure}

\section{ Axial Contribution to the $t\bar{t}$ Total Cross Section
and the Production of Stop-Antistop near Threshold}

The total cross section for the
production of P-wave states near threshold gives the leading term
for the production of stop-antistop ($\t\bar{\t}$) and a contribution
of $O(\a_s^2)$
to the total cross section for $t\bar{t}$. Only qualitative investigations
on the basis of the Coulomb Green function have been
published up to now \pcite{Bigi2,Fad2}. We will apply the method of
replacing divergent nonrelativistic graphs, implicit in the Green
function approach,
by the relativistic, finite ones to get a reliable quantitative
result also for this case.

Considering the Coulomb Green function
as a first approximation it was observed in \pcite{Bigi2} that the total
cross section for a stop-antistop pair near threshold develops an
unphysical linear divergence due to the nonrelativistic approximation.
Even a relativistic calculation of the imaginary part of the one loop
contribution using the commonly used propagator $1/(p^2-m^2+im\G)$
remains logarithmically divergent. This is due to the fact that this
propagator does not fulfill the requirements of (local) quantum field theory
as does the expression
\beg  \plabel{lehm}
 \int \frac{d\m^2}{\pi} \frac{m \G}{(\m^2-m^2)^2+m^2\G^2}
      \frac{1}{p^2-\m^2+i\e},
\ee
which is in agreement with the Lehmann representation of the full
propagator.
The tree contribution (denoted $\s_1$) using \pref{lehm} reads
\beg
 \s_{1}= \frac{3}{2}(c_s+d_s \r) \int d\m_1^2 \int d\m_2^2 \D(\m_1^2)
         \D(\m_2^2)
\Th(P^2-(\m_1-\m_2)^2)\left[1-2\frac{\m_1^2+\m_2^2}{P^2}+
      \frac{(\m_1^2-\m_2^2)^2}{P^4} \right]^{\frac{3}{2}} \s_{\m\bar{\m}},
\ee
whereas we have for the nonrelativistic tree contribution
\beg
\s_{1,nr}^{\L}= \frac{3 \G}{\pi m^2}(c_s+d_s \r) \int_0^{\L} dp
           \frac{p^4}{(p^2-mE)^2+m^2\G^2} \s_{\m\bar{\m}}.
\ee
However, a single subtraction of the tree graph is not sufficient.
One should also subtract the logarithmically divergent one-gluon exchange
term. But since
we replaced the gluon propagator by the resummed one we should also
calculate the relativistic graph with a resummed propagator to get the
correct behavior at infinity. Unfortunately this leads to renormalon
ambiguities. Therefore, we conclude that the uncertainties in choosing
the right cut-off are of the same order of magnitude as higher QCD
corrections, and we circumvent these difficulties for the time being by
keeping the cut-off in the logarithmic divergent terms,
choosing its value $\L=\L_0=m$ from
the observation that this cut-off would have given the correct
answer in the case of the FB-asymmetry discussed in the last section.\\
\begin{figure}
\begin{center}
\leavevmode
\epsfxsize10cm
\epsfbox{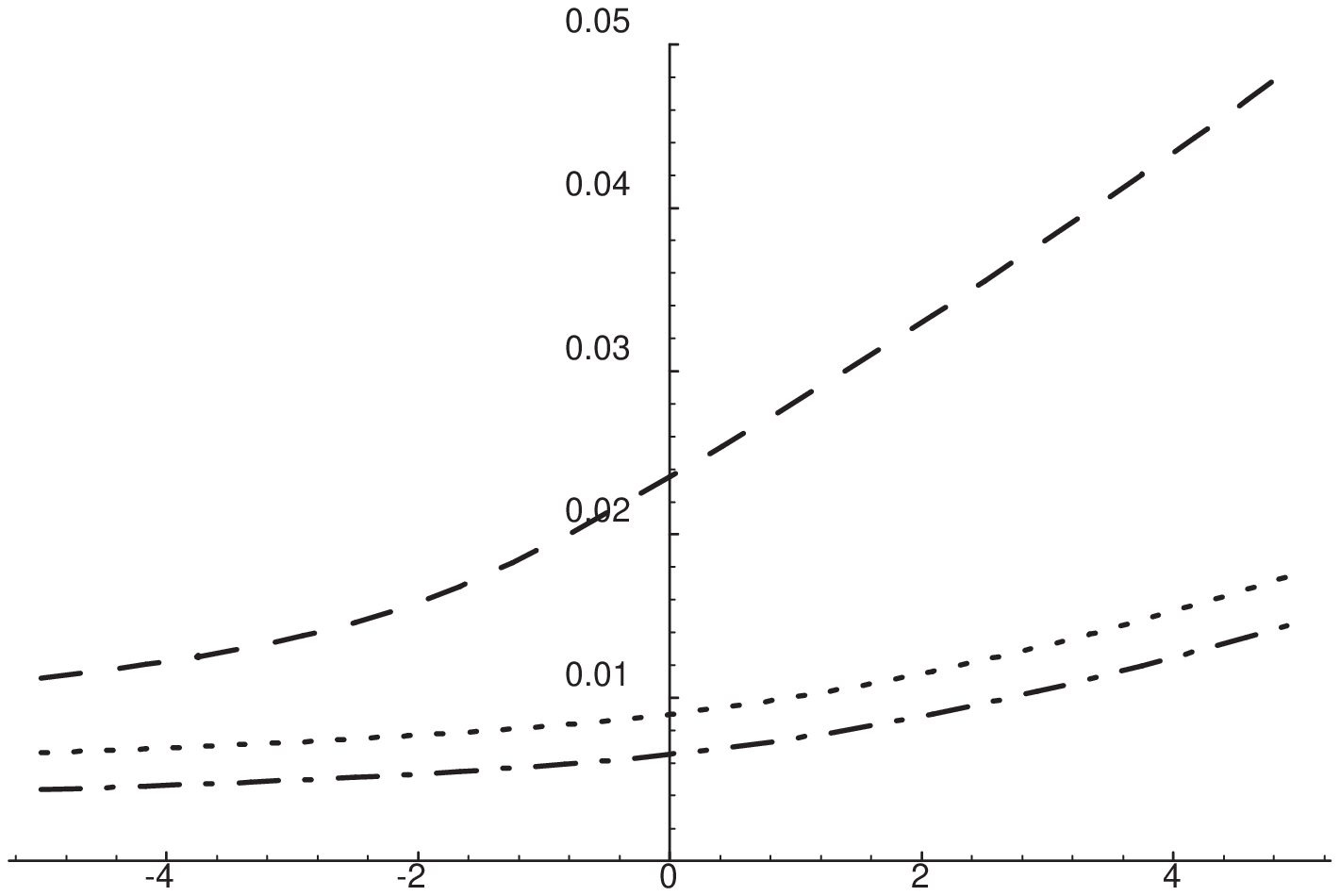}\\
Fig. \refstepcounter{figure} \plabel{stfig1}\thefigure:
Different Contributions to $\s_{\tilde{t}\tilde{\bar{t}}}$\\
dashed:$\s_{nr}^{\L_0}$,dotted:$\s_{1,nr}^{\L_0}$,
dashed-dotted: $\s_{1,rel}$
\end{center}
\unitlength1mm
\begin{picture}(0,0) \put(77,92){$\frac{\s}{\s_{\m\bar{\m}}}$}
\put(130,28){E/GeV}
\end{picture}
\end{figure}
The same is true for the pure axial contribution to the total cross section
for $t\bar{t}$ production since each axial vertex contributes in the
nonrelativistic limit an extra power in $\p$ as explained in the last
section. The only change is to replace the factor
\beg \plabel{fs}
 f_s :=\frac{3}{2}(c_s+d_s \r)
\ee
by
\beg \plabel{ff}
 f_f := 6 (c_{A} +d_{A} \r).
\ee
Using the methods of the preceding sections we obtain within the Green
function approach
\beg
 \s_{nr}^{\L} = f_X \frac{3 \G}{ 2 \pi^2 m^2} \int_0^{\L} dp p^2 \int d\O
\left| \int \frac{d^3q}{(2\pi)^3} \frac{q^{(3)}}{m} G(\q,\p) \right|^2
 \s_{\m\bar{\m}},
\ee
where $f_X$ denotes either  $f_s$ for the scalar or $f_f$ for the
axial fermionic cross section.
The different contributions to the cross section for the scalar case
\beg
 \s = \s_{nr}^{\L_0}-\s_{1,nr}^{\L_0} + \s_1
\ee
are depicted in \fig{stfig1} for $\cos^2 \th_t =
0.5,m=180,\G=\G_{top}$. The net result is shown in \fig{stfig2}
for the axial contribution to the $t\bar{t}$ cross section.

\begin{center}
\leavevmode
\epsfxsize10cm
\epsfbox{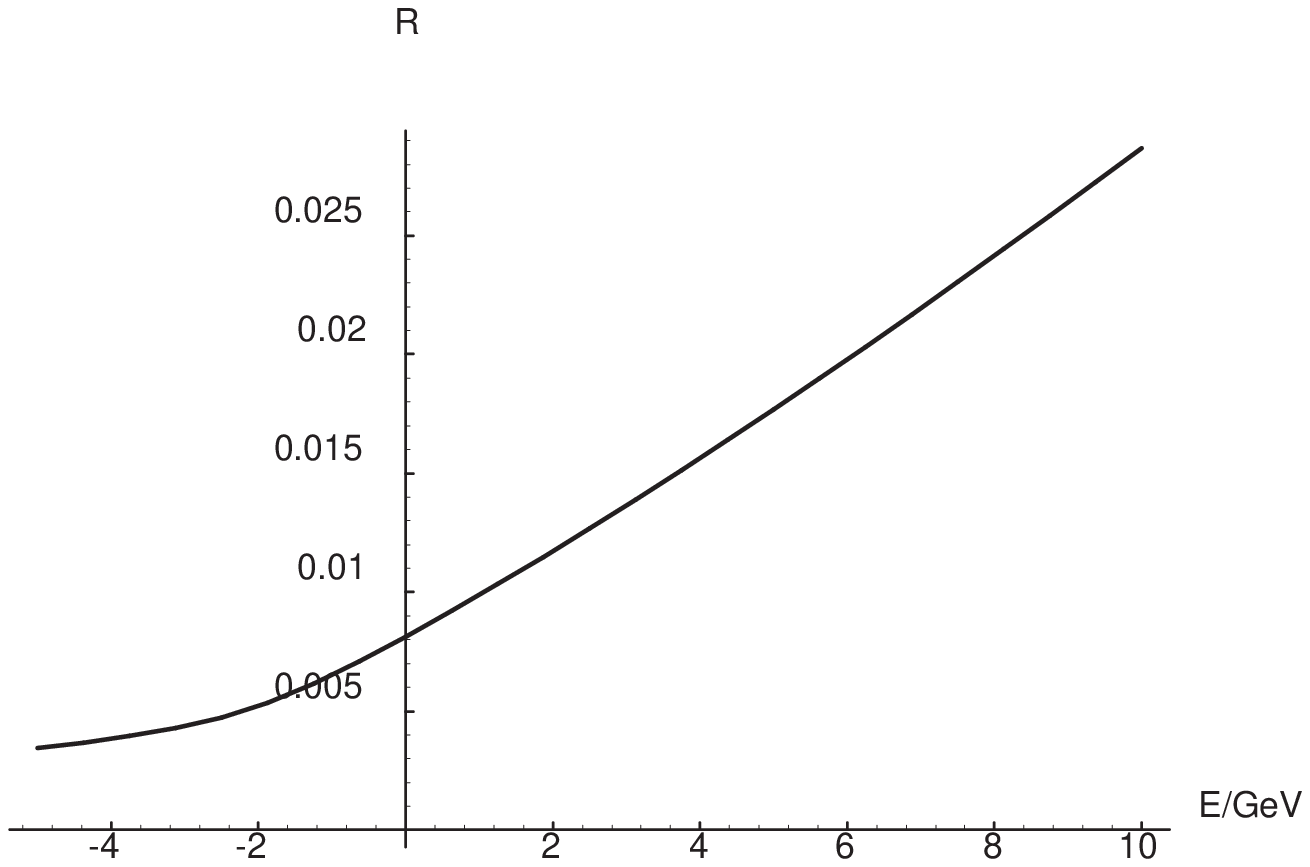}\\
Fig. \refstepcounter{figure} \plabel{stfig2}\thefigure:
Drell-ratio (R$=\s_A/\s_{\m\bar{\m}}$) for the production of $t\bar{t}$
\\
\end{center}

The full $O(\a)$ corrections are taken into account by including the
hard corrections by a factor $(1-\frac{\a}{\pi})^2$  \pcite{Gues} and
a 10\% reduction of the decay width \pcite{kuehn} in addition to the
aforementioned potential. We conclude that the axial contribution
gives a sizeable effect (of a
few percent ) to $\s_{tot}$ beginning at $\approx 5$GeV above threshold.
However, it is small enough in order to hide the uncertainty in $\L$
with respect to $O(\a_s^2)$ corrections to the vector contribution to
$\s_{tot}$.

In \fig{stfig3} we compare the cross sections for the
production of $\t\bar{\t}$ near threshold for two different values of the
decay constant ( $\cos^2 \th_t =0.5,m=180$).

\begin{center}
\leavevmode
\epsfxsize10cm
\epsfbox{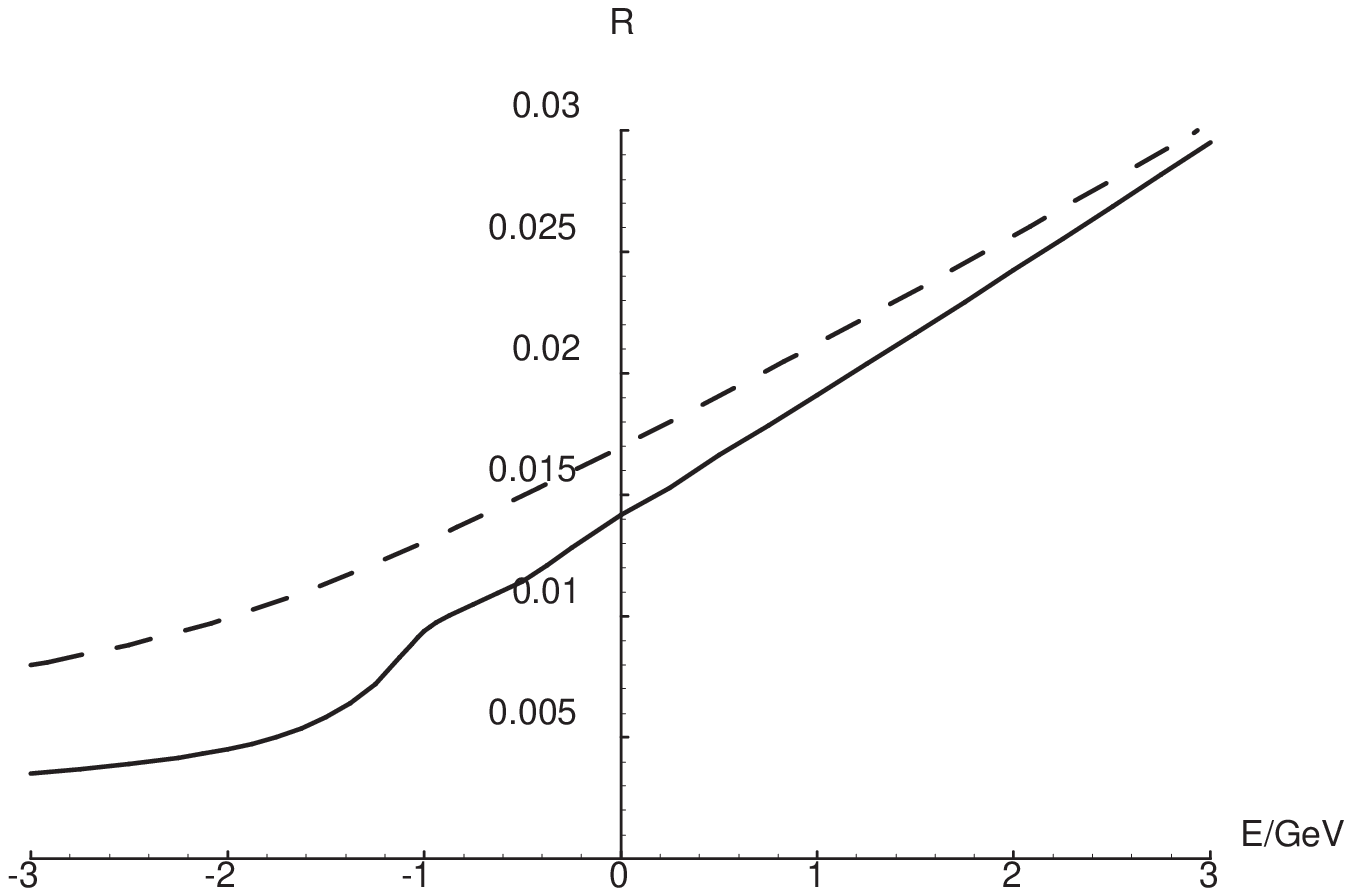}\\
Fig. \refstepcounter{figure} \plabel{stfig3}\thefigure:
dashed curve: $\G=1.5$GeV, full line: $\G=0.5$GeV
\\
\end{center}

It is interesting to note that in contrast to the S-wave case the
$O(\a_s)$ corrections to the potential give a relatively small
effect. This raises the hope that in the P-wave case
fixed order perturbation theory works better as it does for S-waves
and thus our subtraction procedure is stricly applicable also for
the one gluon exchange term without getting troubles from the
resummed gluon propagator.
In \fig{stfig4} and \ref{stfig5} we compare our
result for the $\t\bar{\t}$ cross section with that for a pure Coulomb
potential, both computed with our subtraction method.

\begin{center}
\leavevmode
\epsfxsize10cm
\epsfbox{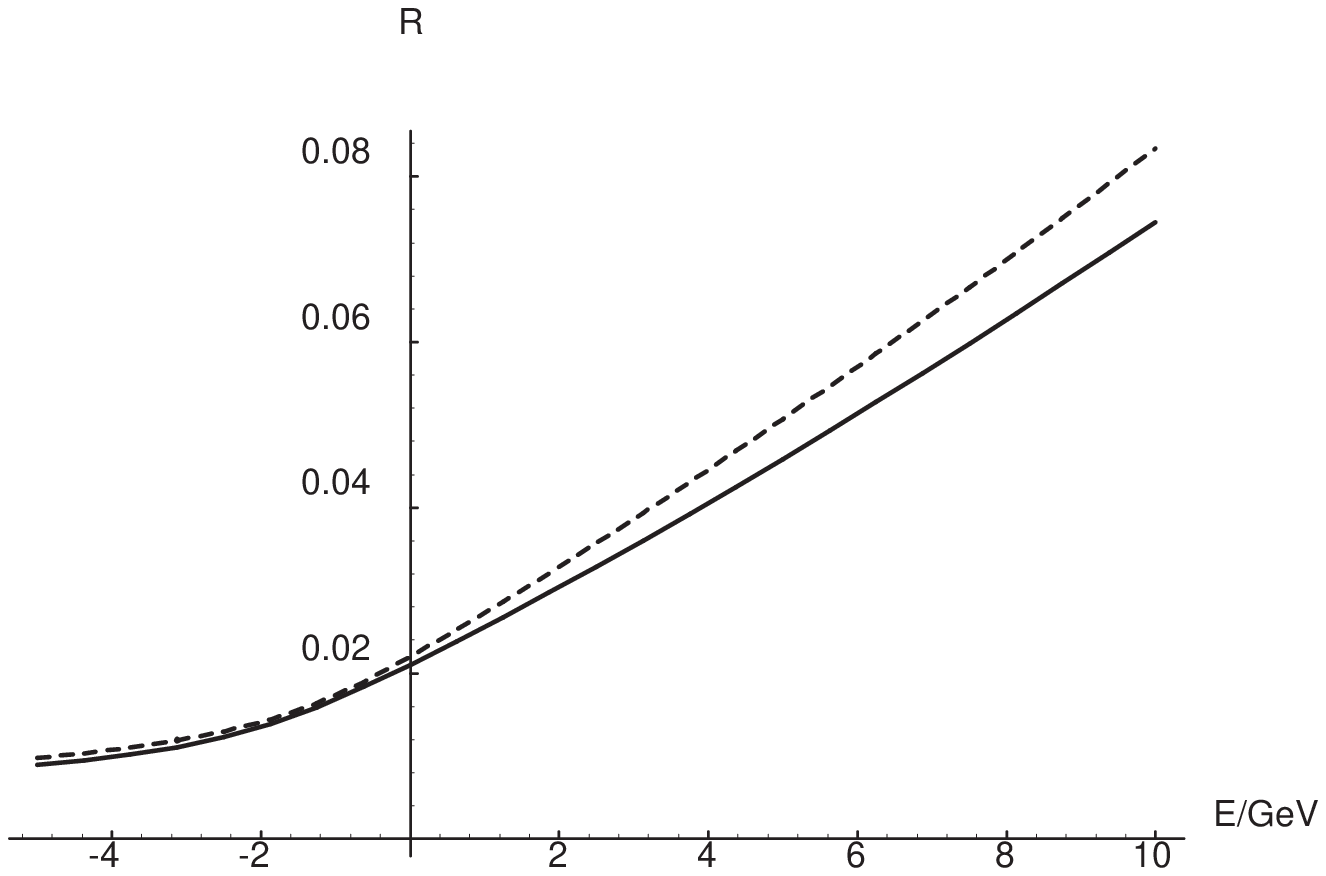}\\
Fig. \refstepcounter{figure} \plabel{stfig4}\thefigure:
dashed curve: Coulomb potential; full line: QCD-potential; both with
$\G=1.5$GeV
\\
\end{center}\begin{center}
\leavevmode
\epsfxsize10cm
\epsfbox{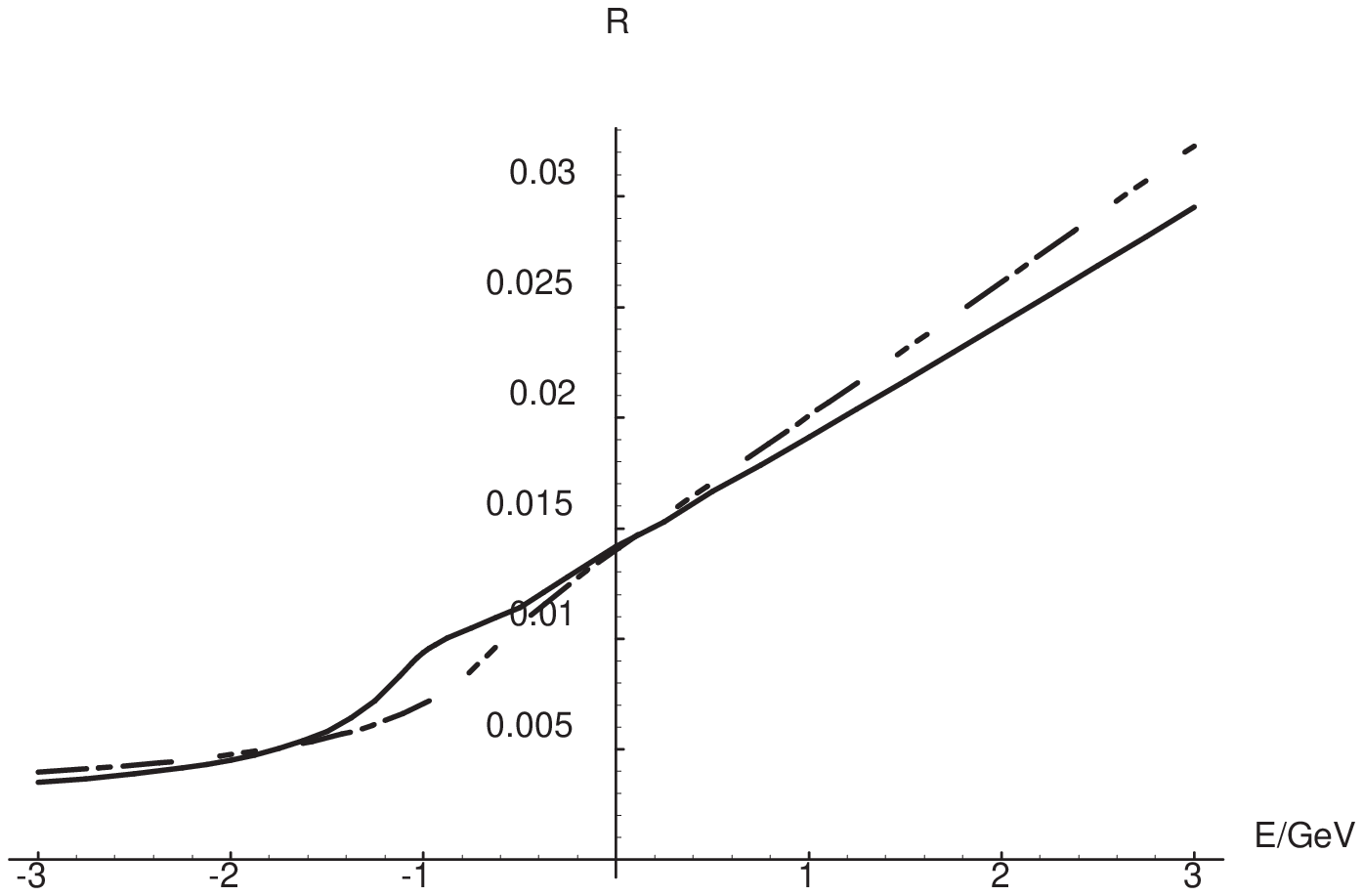}\\
Fig. \refstepcounter{figure} \plabel{stfig5}\thefigure:
dashed curve: Coulomb potential; full line QCD-potential; both with
$\G=0.5$GeV
\\
\end{center}

A further difference is that the lowest lying resonance peak (the 2P-level)
is enhanced by the QCD potential and not suppressed as is the 1S-peak. This
is in agreement with the calculation of the wave function corrections
\pcite{Wirth}.

\section{Conclusion}

We have calculated the forward-backward asymmetry and the pure axial
contribution
to the total cross section for the $t\bar{t}$ system. Furthermore the
leading contribution to the total cross section for heavy scalars is
determined in this region.
Since the necessity of the inclusion of an infinte sum of Coulomb
interactions in practise enforces
the use of a nonrelativistic Schr\"odinger equation, unphysical
divergencies are produced in the nonrelativistic limit.
These are removed by showing that they originate in the leading
nonrelativistic
contributions which may be subtracted and subsequently replaced by the
relativistic expressions which lead to finite results.
Thus we are able to combine both the advantages of a
nonrelativistic calculation near theshold with the one of a strictly
perturbative treatment.

\vspace{1cm}

{\bf Acknowledgement:} This work has been supported by the Austrian Science
Foundation (FWF), project P10063-PHY within the framework of the
EEC- Program
"Human Capital and Mobility", Network "Physics at High Energy Colliders",
contract CHRX-CT93-0357 (DG 12 COMA). I am grateful to
Y.Sumino for sending me the essential parts of his Fortran code
for the numerical evaluation of the Green function. I am also
grateful to M.Je\.zabek, V.A. Khoze and J.H. K\"uhn for instructive
discussions and to W. Kummer for a careful reading of the manuscript
and many useful suggestions.

\small
\renewcommand{\baselinestretch}{1.2} 

\end{document}